\documentclass[journal=jacsat,manuscript=article]{achemso}

\usepackage[version=3]{mhchem} 
 \usepackage[utf8]{inputenc}

\author{Varsha Kumari}
\email{vakumari@iu.edu}
\affiliation{Department of Chemistry, Indiana University, Bloomington, Indiana 47405, United States.}
\author{Julia Bauer}
\affiliation{Department of Applied Physics, Yale University, New Haven, Connecticut 06520, United States.}
\author{Alexandru B. Georgescu}
\email{georgesc@iu.edu}
\affiliation{Department of Chemistry, Indiana University, Bloomington, Indiana 47405, United States.}

\title[An \textsf{achemso} demo]
  {Symmetry-Driven Trimer Formation in Kagome Correlated Electron Materials }

\abbreviations{IR,NMR,UV}

\usepackage{graphicx}
\usepackage{adjustbox}
\begin{document}






\begin{abstract}
Correlated electron materials with molecular orbital states extending over transition metal clusters can host multiferroicity, spin frustration, and unconventional insulating phases. However, the fundamental criteria that govern cluster formation and stability remain unclear. Here, we identify a symmetry, correlation, and electron-filling-driven criteria that stabilize triangular metal trimers in materials displaying transition metal kagome patterns. Using density functional theory and chemical bonding analysis, we show that trimer formation emerges when 6–8 electrons occupy molecular orbitals derived from transition metal d-states, achieving near-complete filling of bonding states while avoiding antibonding occupation, and correlations are of intermediate strength. This principle explains the stability of Nb$_3$X$_8$ (X = Cl, Br, I), and more broadly, our findings offer a general design rule to obtain quantum materials with quantum states extended across transition metal clusters.
\end{abstract}



  

\subsection{Introduction}

\begin{figure}[h]
    \centering
    \includegraphics[height=0.5\textheight]{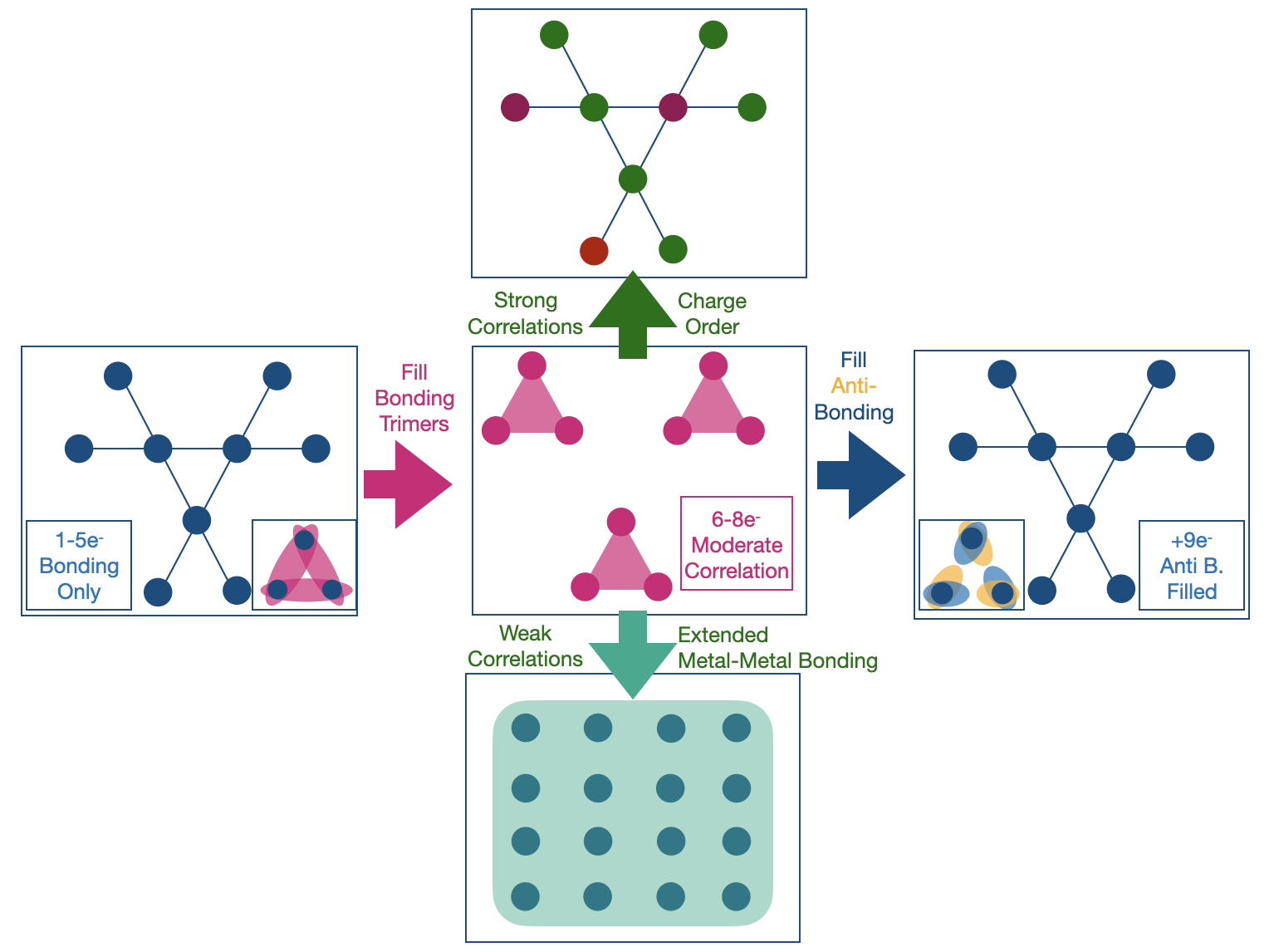}
    \caption{Trimer formation in a Kagome lattice requires an optimum filling of bonding to antibonding molecular orbital states, and an intermediate level of correlation strength on the transition metal ion.}
    \label{fig:Main}
\end{figure}

Correlated electron materials display a wide range of controllable emergent properties, including magnetoresistance~\cite{Helmot1993} and high-temperature superconductivity~\cite{Isakov2006, Wang2013}, enabling applications across microelectronics~\cite{Jiho2020}, bioimaging, information storage, and quantum information technologies~\cite{Morey2019}. These properties arise from the interplay of local states—primarily derived from $d$- and $f$-orbitals—and long-range interactions between them.

In certain materials, local $d$-orbital states form molecular orbital-like states extended across clusters of transition metal ions, leading to the formation of dimers, trimers, and higher-order transition metal clusters; we will refer to these materials more broadly as Correlated Electron Molecular Orbital Materials (CEMO) materials\citep{Attfield, khomskii2020}. 

In this work, by focusing on the example of triangular trimers in a kagome lattice, we show that two key conditions must be met to allow for cluster formation 1) an intermediate level of electron localization/correlation strength on the transition metal ions: strong enough to lead to local molecular orbital formation, and weak enough not to lead to charge order and 2) A just-right electron filling ratio of the bonding/antibonding trimer orbital states - in this case of 6-8 electrons, which is the electron count found in all trimer cluster materials. These criteria are likely general to cluster formation in general. 

Well known examples of materials in this class and cluster formation include the dimerization and metal–metal bond formation in VO$_2$ \cite{zhang2018}, the heptamer network of V–V bonds in AlV$_2$O$_4$~\cite{Horibe2006,Browne2017}, and lacunar spinels\citep{SpinelsHaule}. A key aspect of these materials is that the quantum states are extended across the transition metal clusters, as opposed to the ions themselves. Theoretically, they can also provide a testbed for beyond-DFT theories, as theoretical methods that include correlations at an d-orbital level as opposed to molecular orbital level may fail to reproduce their key electronic features\citep{SpinelsHaule,Rosner,strand}. In other materials, the tendency to form trimers and other clusters may be interconnected with their superconducting behavior\citep{WilsonKagome}.

In the case of trimers, metal–metal bonds and local quantum states are formed on clusters of three transition metal ions. Such cluster formation can arise in a wide variety of materials, including LiVS$_2$, LiVO$_2$~\cite{Kojima2019, D}, TiCrIr$_2$B$_2$~\cite{Gordon2016}, Na$_2$Ti$_3$Cl$_8$~\cite{Meyer}, and KTi$_4$Cl$_{11}$~\cite{Corbett}. Here, we focus on triangular trimers, but we note that some particularly promising frustrated magnets display linear trimers\cite{Ba4Ir3O10}.

\begin{figure}[h]
    \centering
    \includegraphics[height=0.5\textheight]{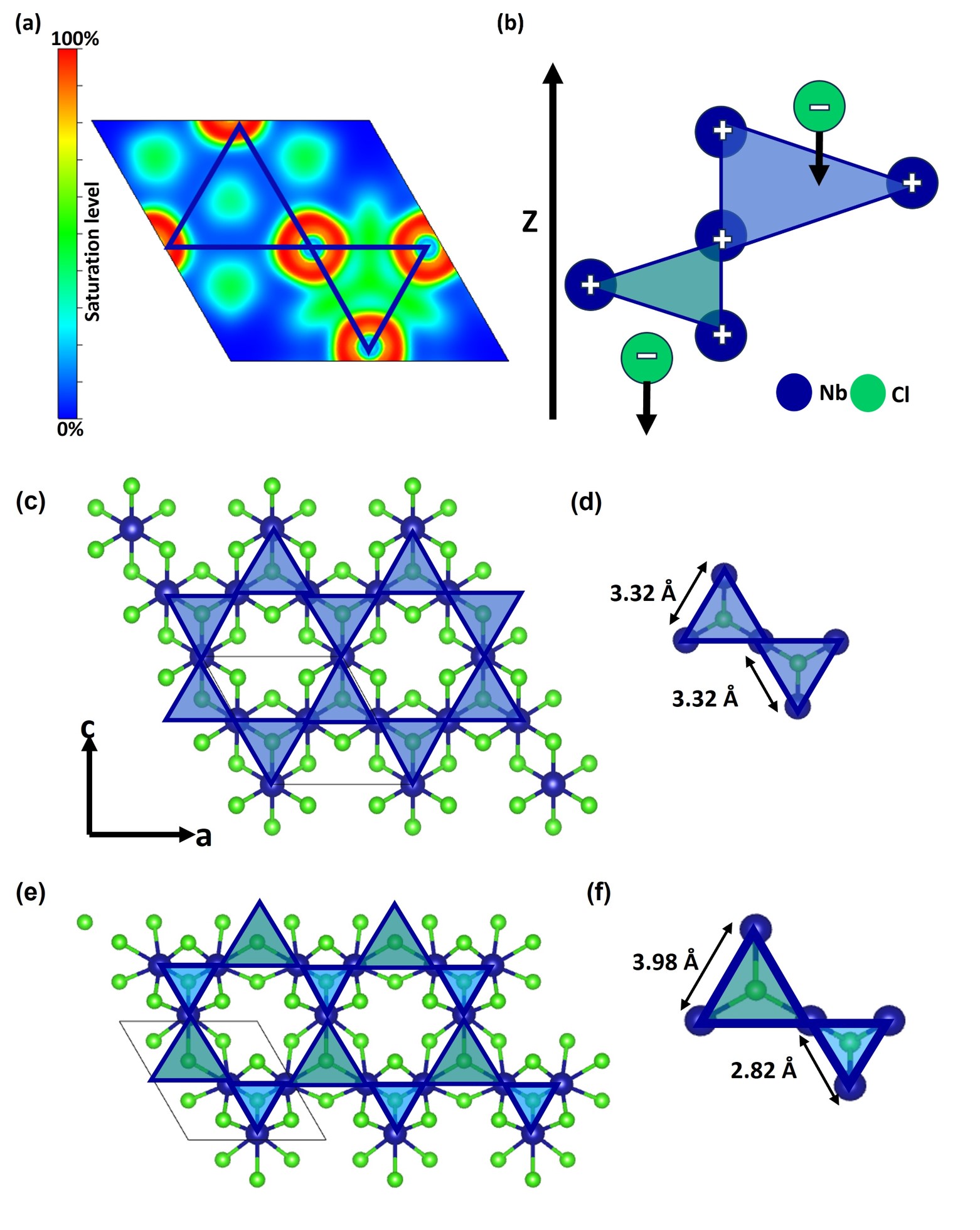}
    \caption{(a) Electron localization function plot showing localization of electrons within a smaller Nb$_3$ cluster. (b) Displacement of apical chlorine along smaller and larger equilateral triangles leads to a polar mode. (c) Crystal structure of Nb$_3$Cl$_8$ (space group $P\bar{3}m1$, no. 164) with spin polarization without breathing distortion (BD). (d) Bond lengths of the smaller and larger equilateral triangles are 2.82~$\AA$ and 3.98~$\AA$, respectively. (e) Crystal structure with spin polarization and BD. (f) Bond lengths of the equilateral triangles are 3.32~$\AA$.}
    \label{Nb3Cl8}
\end{figure}

While metallic, transition metal intermetallic kagome materials do not exhibit trimerization (e.g., FeSn~\cite{MKang2020}, CoSn~\cite{ZLui2020} - or may exhibit it in small amplitudes \citep{WilsonKagome}), ionic, insulating kagome lattices can favor the formation of trimers coupled to a structural breathing distortion of large amplitude, leading to alternating small and large triangles of transition metal ions. The small M$_3$ triangular clusters can subsequently support localized spin states on trimer molecular orbitals; the localized nature of the states can also be seen in the electron localization function (ELF) plot shown in Figure \ref{Nb3Cl8}a. These localized trimer states are arranged into a triangular lattice, potentially giving rise to spin frustration and entanglement~\cite{Khomskii2021}. Such effects, coupled with the unique orbital symmetry of the trimer states, have been proposed to stabilize exotic correlated phases, including possible spin liquid states~\cite{liu2024, Yuya2015}.

The breathing distortion in the kagome layers induces a polar distortion by displacing negatively charged anions, rendering each layer ferroelectric. The coupling between magnetic states localized on trimer orbitals and the polar modes driven by the breathing distortion has led to the classification of these materials as a new class of multiferroics\citep{Wyckoff2019}, as shown in Figure \ref{Nb3Cl8}b . These coupled order parameters have enabled the realization of the first field-free Josephson diode using Nb$_3$Br$_8$, which exhibits superconductivity in one direction and insulating behavior in the opposite direction—an effect that can be used for novel electronic and quantum device applications\cite{Wang2022,Ramesh2007}.

In this paper, we focus on Nb$_3$Cl$_8$ as a model system for exploring a broader family of correlated electron materials based on the structural unit [M$_3$X$_8$], where M is a transition metal and X is a ligand. These compounds feature partially filled d-orbitals and moderate electron correlations. The Nb$_3$X$_8$ series (X = Cl, Br, I) forms two-dimensional layered structures held together by van der Waals forces, and can be exfoliated for experimental investigations, where the number of layers is precisely controlled\cite{grytsiuk2024,Haraguchi}; for the purposes of our calculations, this also makes them an ideal model system.

Unlike other Kagome materials displaying trimerization, the Nb$_3$X$_8$ series is unique: others take the form of M$'_{x}$M$_3$X$_8$, and depend on interstitial species for stabilization\citep{NaKagome}. We hypothesize that this distinction is closely linked to the intrinsic stability of cluster formation in the Nb$_3$X$_8$ system.

To elucidate the origin of the unique stability of Nb$_3$X$_8$ compounds, and of the design parameters of trimer formation in kagome materials - and transition metal clusters in solids more broadly -, we perform density functional theory (DFT) and DFT+$U$ calculations, which reveal a strong interplay between breathing distortions (BD), trimer orbital formation, and electronic correlation strength. In these materials, Nb atoms are arranged in a kagome lattice, and a structural breathing distortion drives the formation of Nb$_3$ triangular clusters. These clusters host trimer molecular orbitals whose symmetry and electron filling are optimized by the BD, leading to near-ideal occupation of bonding and antibonding states around the Fermi level, and an intermediate level of electronic correlation (Fig \ref{fig:Main}). This electronic configuration improves the chemical stability of the system. Using Nb$_3$Cl$_8$ as a model compound, we systematically disentangle the roles of BD, trimerization, and correlation effects. The insights obtained are likely broadly applicable to other transition metal clusters in a variety of coordinations.

\section{Results and Discussion}

We use Nb$_3$Cl$_8$ as a model system to explore the interplay between spin polarization and breathing distortions in governing the relationship between structural symmetry and chemical stability. In addition, we systematically investigate the effects of metal and anion substitution, along with variations in the electronic correlation strength, to understand how symmetry breaking and electronic interactions collectively stabilize - or destabilize - the trimerized kagome structure.

\subsection{Electronic Structure of Nb$_3$Cl$_8$}

DFT calculations of a high symmetry structure of Nb$_3$Cl$_8$ in which the Nb atoms form an ideal kagome lattice without a breathing distortion (BD), as shown in Figure \ref{Nb3Cl8}c. In this undistorted configuration, all Nb–Nb bond lengths are equal, forming equilateral triangles with a bond length of 3.32~\AA\ (Figure \ref{Nb3Cl8}d), and trimer molecular orbitals do not form. When symmetry breaking is allowed, the DFT-optimized structure (Figure \ref{Nb3Cl8}e) closely agrees with experimental data and exhibits a clear BD, consisting of alternating small and large Nb$_3$ triangles with bond lengths of 2.82~\AA\ and 3.98~\AA, respectively (Figure \ref{Nb3Cl8}f) - in good agreement with experiment (2.82~\AA\ and 3.92~\AA)\citep{experiment}. This distortion favors the formation of trimer molecular orbitals along the Nb–Nb bonds. Within non-magnetic DFT, and in the absence of a Hubbard $U$ correction, the structure with a breathing distortion is energetically favored by 181~meV/atom over the undistorted one. 

DFT calculations show that the energy difference between the spin-polarized and non-spin-polarized ground states of Nb$_3$Cl$_8$ is only 4~meV/atom, indicating that magnetic ordering contributes minimally to the material’s overall stability. In contrast, the breathing distortion (BD) plays a much more significant role in stabilizing the structure, as previously discussed.

\begin{figure}[h]
    \centering
    \includegraphics[width=1\linewidth]{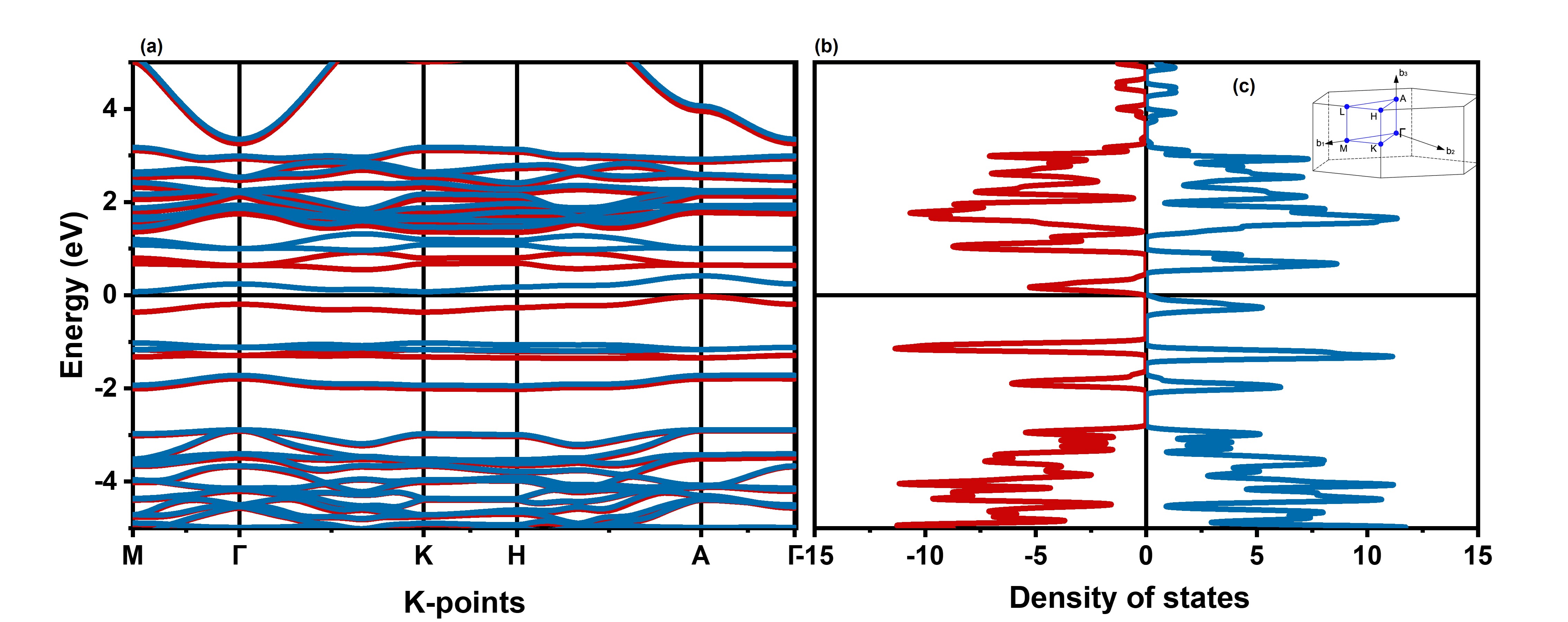}
    \caption{(a) Electronic band structure and (b) Spin-polarized density of states (DOS) of Nb$_3$Cl$_8$ obtained from DFT calculations.}
    \label{Band & DOS}
\end{figure}

The band structure and density of states (DOS) for the non-magnetic, spin-polarized Nb$_3$Cl$_8$ structure with BD (Figures \ref{Band & DOS}a and \ref{Band & DOS}b) show narrow, flat bands near the Fermi level. These features point to strong electron localization within the Nb$_3$ clusters and the formation of trimer molecular orbitals. The molecular orbital localization and narrow bandwidth make it likely this is in a Mott regime, where strong electronic correlations dominate\citep{Rosner,strand}.

\begin{figure}[h]
    \centering

    \includegraphics[height=0.25\textheight]{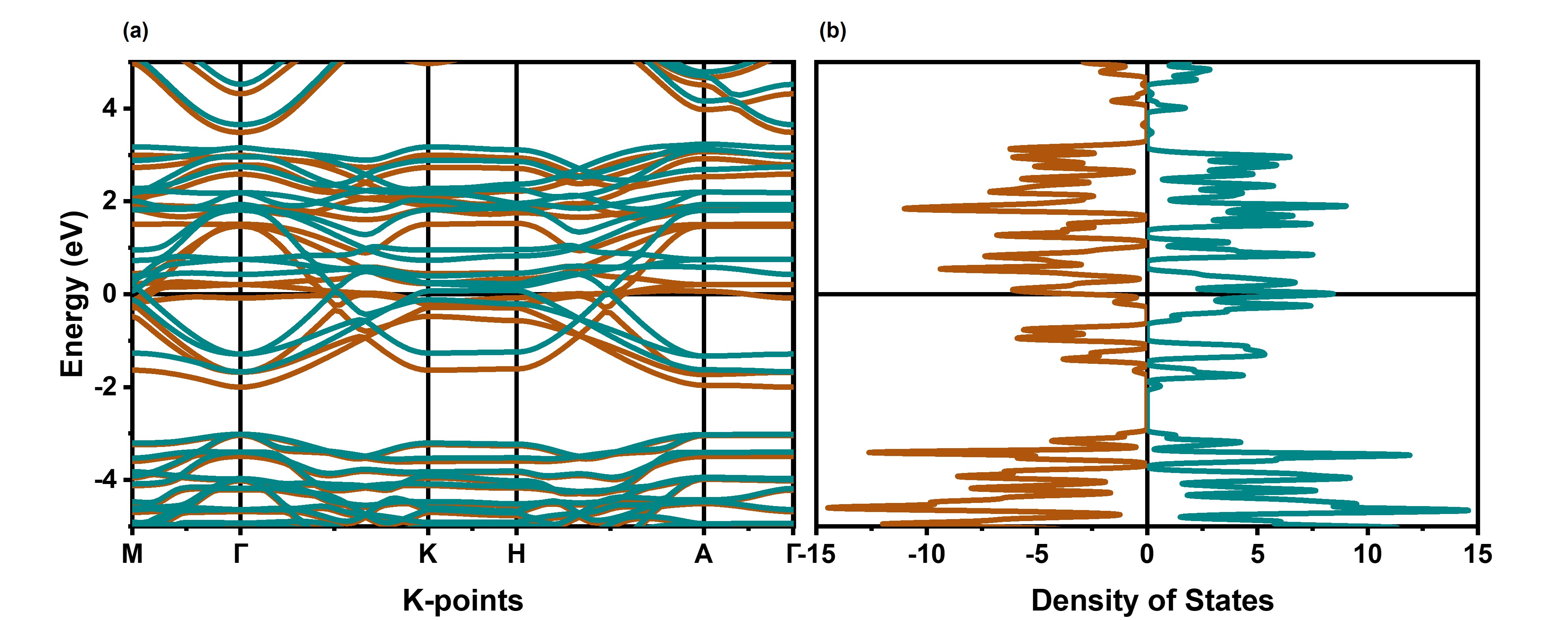}
    \caption{(a) Electronic band structure and (b) DOS of the ideal Kagome lattice of Nb$_3$Cl$_8$ obtained from DFT calculation.}
   \label{Bands}
\end{figure}

In contrast, the electronically active states in the ideal kagome lattice of Nb$_3$Cl$_8$—without breathing distortion—are relatively delocalized. As shown in Supporting Information Figures \ref{Bands}a and \ref{Bands}b, the $d$-orbital-derived bands near the Fermi level span a bandwidth of approximately 1~eV, indicative of more itinerant electronic behavior. 

\subsection{Electronic Origin of Trimer Stability}

\begin{figure}[h]
    \centering
    \includegraphics[width=1\linewidth]{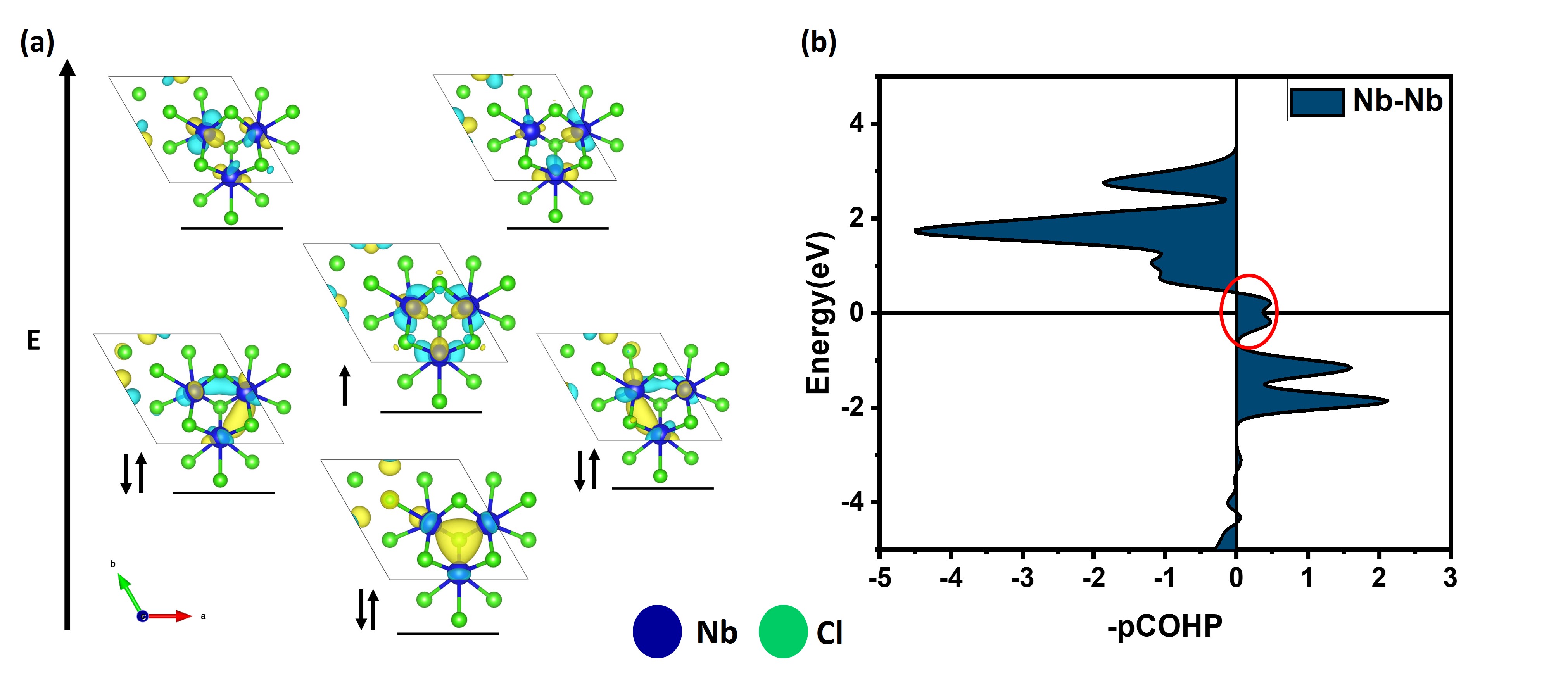}
    \caption{(a) Energy levels of Nb$_3$ cluster plotted at $\Gamma$ point (b) COHP plot of Nb$_3$Cl$_8$ showing optimal filling of bonding to anti-bonding trimer orbital near Fermi energy.}
    \label{COHP}
\end{figure}

We visualize selected Bloch wavefunctions near the Fermi level at the $\Gamma$ point, Figure \ref{COHP}a - these wavefunctions are real-valued and allow for direct visualization of orbital symmetry and phase.

Understanding the bonding character of these states is critical, as the occupation of bonding versus antibonding orbitals directly influences the chemical stability of the material. In general, filling bonding orbitals stabilizes a system, while occupation of antibonding orbitals has a destabilizing effect; this is in analogy to, for example, the H$_2$ molecule, which is stable due to its filled bonding orbital, as opposed to the unstable He$_2$ molecule, where equal occupation of bonding and antibonding orbitals results in no net bonding interaction.

The formation and stability of the Nb$_3$ trimers is strongly connected to their electron filling, with seven electrons per trimer playing a key role in governing the material's structural, electronic, and magnetic properties. Each niobium atom contributes five valence electrons, yielding a total of 15 electrons per Nb$_3$ unit. After accounting for charge transfer to the surrounding Cl$^-$ ligands, seven $d$-electrons remain localized within the trimer, consistent with a formal [Nb$_3$]$^{8+}$ charge state.

Electronic structure analysis reveals that the three lowest-energy trimer molecular orbitals are fully occupied, while the fourth is partially filled. This configuration gives rise to a spin-$\frac{1}{2}$ magnetic moment per trimer\cite{Wyckoff2019}.

\begin{figure}[h]
    \centering
    \includegraphics[height=0.45\textheight]{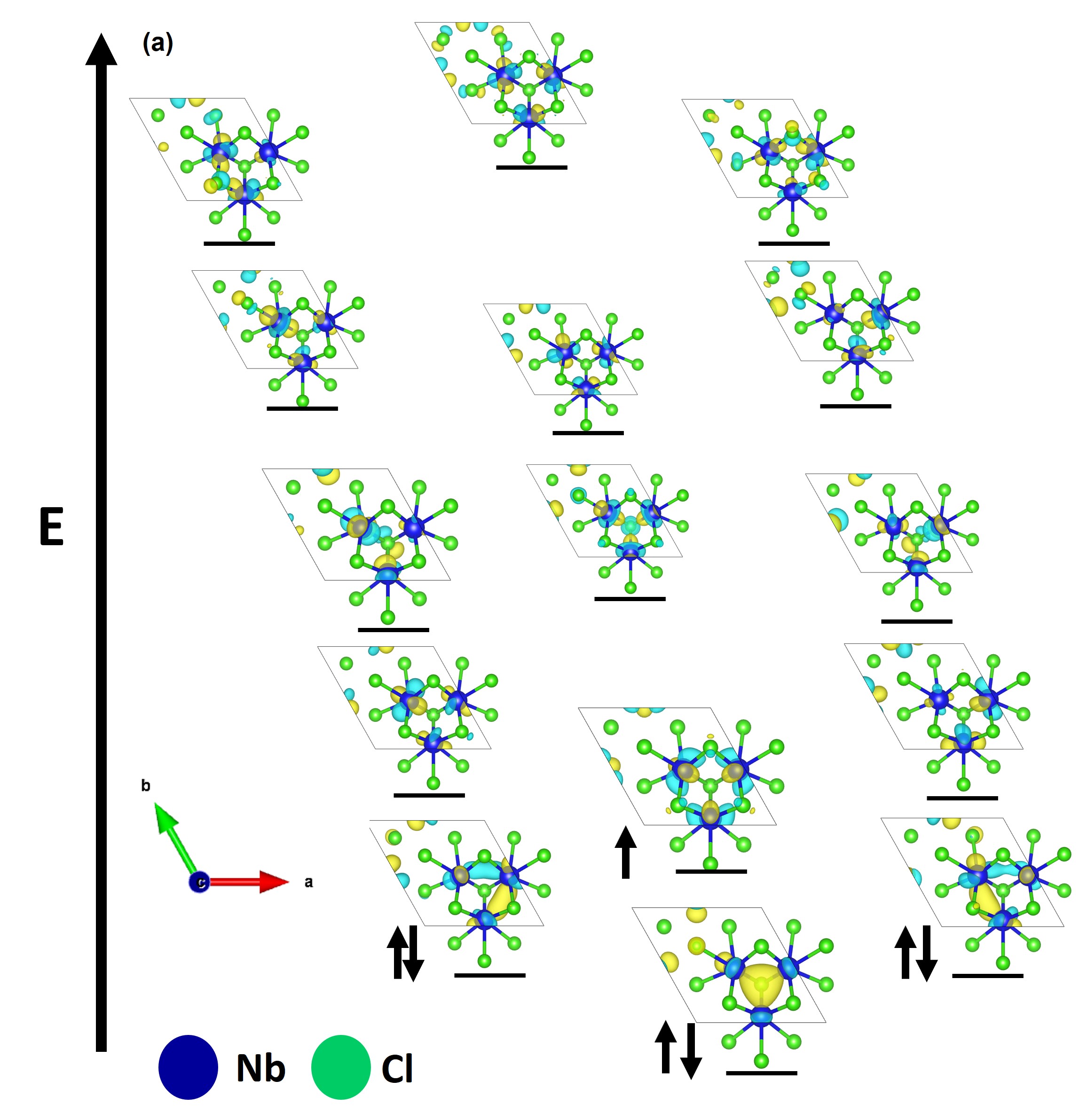}
    \caption{Energy level splitting diagram of 15 trimer orbitals formed from the $d$-orbitals of each Nb atom in the Nb$_3$ cluster.}
    \label{trimer-splitting}
\end{figure}

To quantitatively assess the bonding and antibonding character of the trimer orbitals, we employed the LOBSTER software (Version 5.0) to perform Crystal Orbital Hamilton Population (COHP) analysis. The resulting projected COHP (pCOHP) plot for Nb$_3$Cl$_8$, shown in Figure \ref{COHP}b, provides insight into the relative occupancy of the bonding and antibonding states, which in turn is directly related to the material’s chemical stability, and tendency to form transition metal clusters.

In the pCOHP representation, the bonding interactions (which stabilize the structure) appear on the negative $x$ axis (left side), while the antibonding interactions (which destabilize it) appear on the positive $x$-axis (right side)\cite{Steinberg2018}. The vertical line at zero separates the bonding from the antibonding character, and the horizontal line at 0~eV marks the Fermi level. The Fermi level is set to zero on the y-axis. Our analysis focuses specifically on nearest-neighbor Nb–Nb interactions within the Nb$_3$ cluster.

As shown in Figure \ref{COHP}b, the bonding states near the Fermi level are nearly fully occupied, while the antibonding states remain largely unoccupied. This favorable bonding–antibonding filling configuration contributes significantly to the chemical stability of Nb$_3$Cl$_8$. The red circle highlights a characteristic feature of the spin-polarized ground state—a split peak structure resulting from spin-dependent occupancy of trimer orbitals (although the COHP plot does not resolve spin channels). These results support the conclusion that an optimal electron filling of 6–8 electrons per Nb$_3$ cluster maximizes bonding occupancy while minimizing antibonding contributions. Deviations from this range are expected to destabilize the trimer framework and reduce material stability.

This bonding–antibonding filling principle provides a unifying framework for understanding the stability of the broader Nb$_3$X$_8$ family, where variations in halogen species preserve the electron count and maintain favorable trimer orbital occupancy.

Nb$_3$Cl$_8$ crystallizes in the trigonal space group $P\overline{3}m1$ (No.~164), with each Nb atom occupying a site of $C_{1h}$ point group symmetry. This low site symmetry leads to no degeneracy among the five $d$-orbitals on each Nb atom. Each type of orbital forms three molecular orbitals with identical orbitals on the nearby transition metal ions, leading to the formation of molecular orbitals within the Nb$_3$ trimer. The resulting molecular orbital energy level diagram, comprises 15 trimer orbitals derived from the Nb $d$-orbitals, is shown in Figure \ref{trimer-splitting}.

The formation and splitting of these trimer orbitals can be rationalized using group theory, specifically through symmetry-adapted linear combinations (SALCs). Treating the Nb$_3$ unit as a cyclic system with $C_3$ symmetry (triangular point group) provides a convenient framework for constructing SALCs. Under this symmetry, combinations of atomic orbitals transform as irreducible representations of the $C_3$ group, leading to one non-degenerate ($a_1$) and two degenerate ($e$) orbital sets for each type of atomic orbital. As a result, each set of three atomic $d$-orbitals on the trimer splits into one $a_1$ and a pair of doubly degenerate $e$ orbitals, yielding the characteristic 1–2 pattern of molecular orbital levels observed in Figure \ref{trimer-splitting}.

This symmetry-driven orbital framework is key to understanding the electronic structure, bonding, and stability of the Nb$_3$ cluster and, by extension, the broader Nb$_3$X$_8$ family.

Combined with the 7-electron filling of the Nb$_3$ cluster and supported by COHP analysis, these results confirm that the material’s stability is closely tied to the near-complete occupation of bonding trimer orbitals and minimal occupation of antibonding states.

\begin{figure}[h]
    \centering
    \includegraphics[width=1\linewidth]{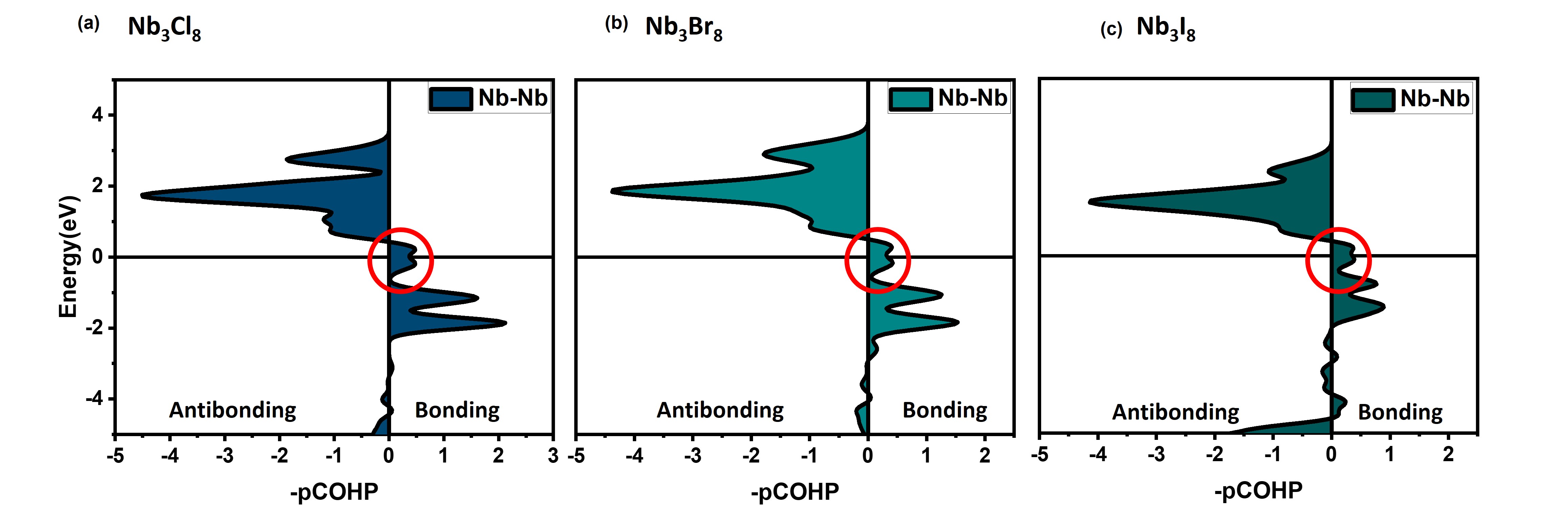}
    \caption{(a) COHP plot of Nb$_3$Cl$_8$ (b) COHP plot of Nb$_3$Br$_8$ (c) COHP plot of Nb$_3$I$_8$ showing optimal filling of bonding to anti-bonding trimer orbital near Fermi energy.}
    \label{COHPplot}
\end{figure}

\subsection{Role of Halogen Substitution in Nb$_3$X$_8$ and Doping in the Nb$_3$Cl$_8$ System}

As discussed in the previous section, COHP analysis underlines the role of optimal bonding/antibonding MO filling Figure \ref{COHPplot}.

Among the layered halides studied, only Nb$_3$X$_8$ compounds with X = Cl, Br, or I have been successfully synthesized as two-dimensional materials. We propose that this synthetic selectivity arises from optimal filling of bonding and antibonding trimer molecular orbitals, an idea supported by the COHP plots shown in Figure \ref{COHPplot}. 

Despite the differences in electronic correlation strength among Cl, Br, and I, as reflected by the systematic increase in the short and long Nb–Nb bond lengths from 2.82$~\AA$
and 3.98$~\AA$ in Nb$_3$Cl$_8$, to 2.89$~\AA$ and 4.26$~\AA$ in Nb$_3$Br$_8$, and further to 3.01$~\AA$ and 4.66$~\AA$ in Nb$_3$I$_8$, all three compounds exhibit the same formal valence state of [Nb$_3$]$^{+8}$ corresponding to a total of seven $d$-electrons per Nb$_3$ unit. The COHP analysis reveals that this consistent electron count maintains favorable bonding–antibonding orbital filling across the series.

These results suggest that the chemical stability of the Nb$_3$X$_8$ halides is primarily dictated by the electronic structure of the Nb$_3$ cluster, rather than the identity of the halogen, provided the 7-electron filling is preserved, with only a minor role played by the relative tuning of correlation strength by halogen substitution. 

\subsection{The Role of Electron Doping and Correlation Strength: Doping the Metal Site}

\begin{figure}[h]
    \centering
    \includegraphics[height=0.35\textheight]{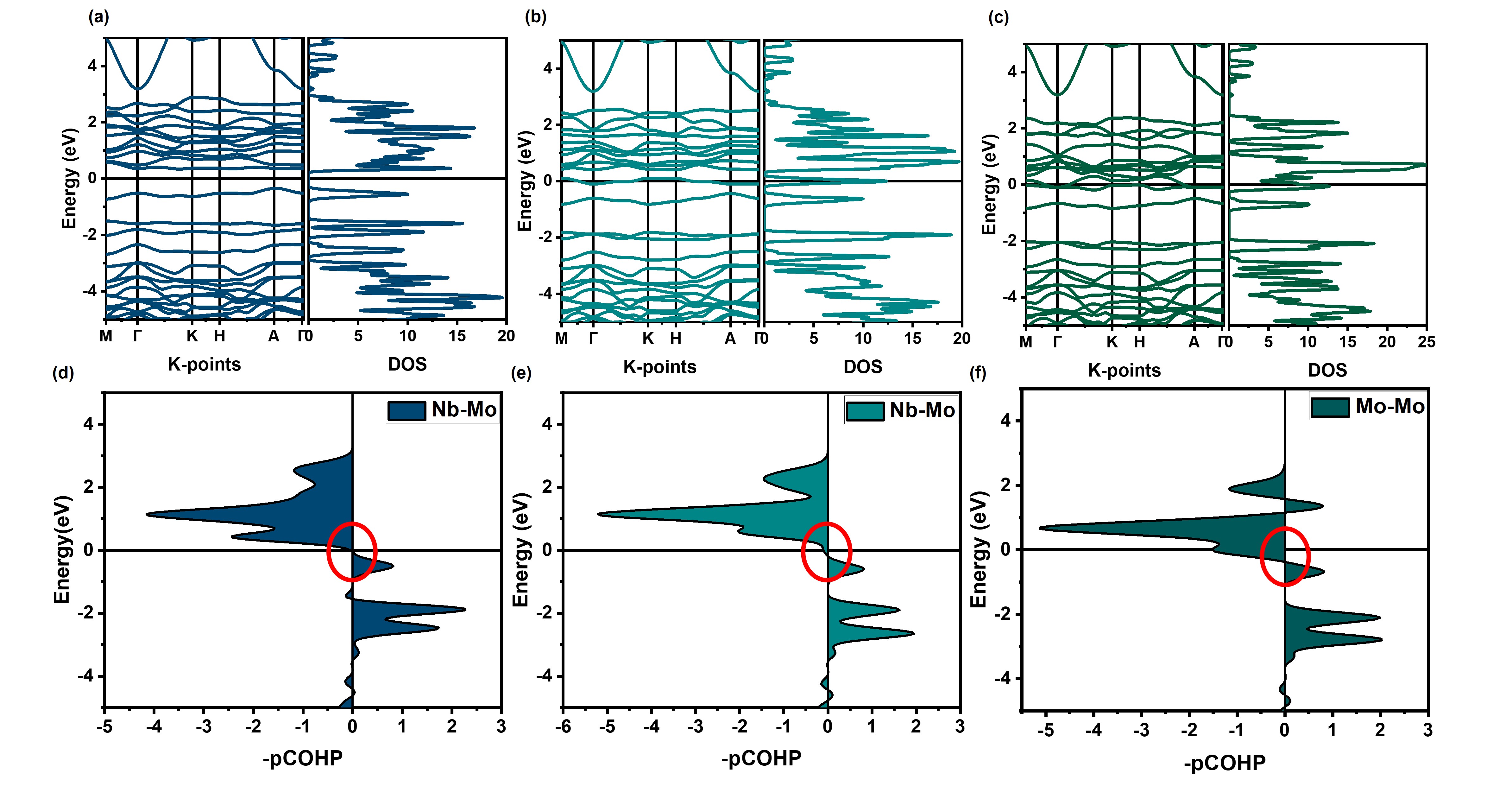}
    \caption{(a) Band structure and DOS of Nb$_2$MoCl$_8$; (b) Band structure and DOS of NbMo$_2$Cl$_8$; (c) Band structure and DOS of Mo$_3$Cl$_8$; (d) COHP plot of Nb$_2$MoCl$_8$ showing complete filling of bonding orbitals near the Fermi energy; (e) COHP plot of NbMo$_2$Cl$_8$ showing filling of anti-bonding orbitals near the Fermi energy; (f) COHP plot of Mo$_3$Cl$_8$ showing significant filling of anti-bonding orbitals near the Fermi energy.}
    \label{Moly}
\end{figure}

Building on the conclusion from the previous section—that the bonding-to-antibonding orbital filling ratio plays a key role in stabilizing kagome halides—we now examine the effects of electron doping via metal substitution. Specifically, we consider doping Nb with a group 6 element such as Mo (molybdenum), which introduces an additional $d$-electron per substitution. This extra electron fully populates the bonding trimer orbitals near the Fermi level, as illustrated in Figure \ref{Moly}. In contrast, doping Nb with Ta (a group 5 element) maintains the total electron count at seven per trimer, enabling an investigation of how reduced electronic correlation strength alone affects the material’s stability.

The incorporation of two different metal species into the M$_3$Cl$_8$ framework—forming M$_{3-x}$M$'_{x}$Cl$_8$ compounds breaks the original equilateral symmetry and introduces distinct short and long bond lengths within the triangular units, as summarized in Table \ref{tab:bond_lengths}. The short Nb–Mo bond length is approximately 2.82$~\AA$, closely matching the sum of their atomic radii. Similarly, the Nb–Ta short bond length measures around 2.75$~\AA$. For comparison, the Nb–Nb bond length is approximately 2.74$~\AA$, which corresponds to twice the atomic radius of Nb. These values fall within the typical range of metal–metal (M–M) bonding and suggest that trimer formation remains favorable in doped systems - provided the number of electrons remains the same - further contributing to their structural stability.

\begin{table}[h!]
    \centering
    \caption{Metal–metal bond lengths (\AA) in M$_{3-x}$M$'_{x}$Cl$_8$ clusters (M$'$ = Mo, Ta). Each row lists two distinct short and long bond lengths corresponding to different atomic pairs.}
    \label{tab:bond_lengths}
    \begin{tabular}{lcccc}
        \hline
        \textbf{Structure} & \textbf{Short 1} & \textbf{Short 2} & \textbf{Long 1} & \textbf{Long 2} \\
        \hline
        Nb$_2$MoCl$_8$ & 2.79 (Nb–Nb) & 2.73 (Nb–Mo) & 3.98 (Nb–Nb) & 3.97 (Nb–Mo) \\
        NbMo$_2$Cl$_8$ & 2.74 (Nb–Mo) & 2.82 (Mo–Mo) & 3.96 (Nb–Mo) & 3.90 (Mo–Mo) \\
        Nb$_2$TaCl$_8$ & 2.80 (Nb–Nb) & 2.81 (Nb–Ta) & 3.95 (Nb–Nb) & 3.95 (Nb–Ta) \\
        NbTa$_2$Cl$_8$ & 2.79 (Nb–Ta) & 2.80 (Ta–Ta) & 3.93 (Nb–Ta) & 3.92 (Ta–Ta) \\
        \hline
    \end{tabular}
\end{table}

To further explore the role of electron filling in stabilizing kagome halides, we investigated the effects of sequential Mo substitution in the Nb$_3$Cl$_8$ system. Hypothetical compounds Nb$_2$MoCl$_8$, NbMo$_2$Cl$_8$, and Mo$_3$Cl$_8$ were simulated to evaluate their electronic structure and chemical stability.

As shown in Figure \ref{Moly}a, Nb$_2$MoCl$_8$ exhibits a flat band and a clear bandgap near the Fermi level, indicating strong molecular orbital localization and an insulating ground state. The corresponding COHP plot (Figure \ref{Moly}d) confirms complete filling of the bonding trimer orbitals. With eight $d$-electrons per cluster, the Nb$_2$Mo unit adopts a formal [Nb$_2$Mo]$^{3+}$ valence, achieving optimal bonding–antibonding orbital occupation. These results suggest that Nb$_2$MoCl$_8$ may be a chemically stable material in the Mo-doped series and may be a viable candidate for experimental synthesis.

In contrast, NbMo$_2$Cl$_8$ and Mo$_3$Cl$_8$ display reduced orbital localization, as seen in Figures \ref{Moly}b and \ref{Moly}c. The COHP plots (Figures \ref{Moly}e and \ref{Moly}f) show that the Fermi level shifts into the antibonding states, resulting in electronic configurations that destabilize the trimer framework. This progressive filling of antibonding orbitals likely correlates with decreased structural stability across the series.

For comparison, Ta substitution in Nb$_3$Cl$_8$ maintains the total electron count at seven per trimer but slightly reduces electronic correlation strength. As a result, the trimer orbital filling remains favorable, and stability is largely preserved, in contrast to the Mo-rich compounds.

In conclusion, Mo doping promotes full occupation of the bonding trimer orbitals, thereby maximizing the bonding–antibonding orbital filling ratio and enhancing structural stability. In contrast, 1/3 and 2/3 Ta-substituted Nb$_3$Cl$_8$ compounds maintain the ideal electron count of seven per M$_3$ unit while slightly reducing electronic correlation strength.

To further investigate the relationship between structural distortion, orbital filling, and material stability, we extended our study to a set of hypothetical M$_3$Cl$_8$ systems: Ti$_3$Cl$_8$, V$_3$Cl$_8$, Cr$_3$Cl$_8$, Zr$_3$Cl$_8$, Ta$_3$Cl$_8$, and Mo$_3$Cl$_8$. These were analyzed to identify trends linking electronic configuration with breathing distortion (BD), trimerization, and structural parameters.

The relaxed structures of these compounds exhibit alternating long and short M–M bond lengths, forming distorted triangles that signal the presence of BD and trimer orbital formation. These bond lengths, along with their ratios $R = L/S$ (long-to-short), are summarized in Table \ref{tab:compact_table}. The table also includes the total number of $d$-electrons per M$_3$ unit and the calculated metal diameter (twice the atomic radius), which closely matches the short bond length. Together, these quantities provide a quantitative basis for correlating electron filling with tendencies for trimer formation, and materials' stability across the series.

\begin{table}[ht]
    \centering
    \resizebox{\textwidth}{!}{%
    \begin{tabular}{|c|c|c|c|c|c|}
    \hline
    \textbf{Structure} & \textbf{e$^-$/M$_3$} & \textbf{M Diameter ($\AA$) } & \textbf{Short ($\AA$) } & \textbf{Long ($\AA$)} & \textbf{R(L/S)} \\
    \hline
    Ti$_3$Cl$_8$  & 4  & 2.72 & 3.03 & 3.72 & 1.22 \\
    V$_3$Cl$_8$   & 7  & 2.50 & 2.68 & 3.76 & 1.40 \\
    Cr$_3$Cl$_8$  & 10 & 2.54 & 2.68 & 3.65 & 1.36 \\
    Zr$_3$Cl$_8$  & 4  & 2.96 & 3.17 & 3.87 & 1.22 \\
    Nb$_3$Cl$_8$  & 7  & 2.74 & 2.82 & 3.98 & 1.41 \\
    Mo$_3$Cl$_8$  & 10 & 2.90 & 2.78 & 3.90 & 1.40 \\
    Ta$_3$Cl$_8$  & 7  & 2.76 & 2.78 & 3.90 & 1.40 \\
    \hline
    \end{tabular}%
    }
    \caption{Structural parameters of M$_3$Cl$_8$ clusters: total electrons per cluster, metal radius, long and short bond lengths, and their ratio R = L/S.}
    \label{tab:compact_table}
\end{table}

To broaden the scope of our study and generalize the design principles for trimer-based kagome materials, we investigate the stability of a broader class of multiferroic insulating compounds. These include materials with structural motifs of the form M$_3$X$_8$ and M$'_y$M$_z$X$_8$, where M is a transition metal, M$'$ is an alkali or alkaline earth element, and X is a halogen or oxygen.\cite{Matthew, Wyckoff2022} This extended set encompasses both known materials and new hypothetical systems with kagome lattices formed by trimer orbitals.

\begin{figure}[h]
    \centering
    \includegraphics[width=1\linewidth]{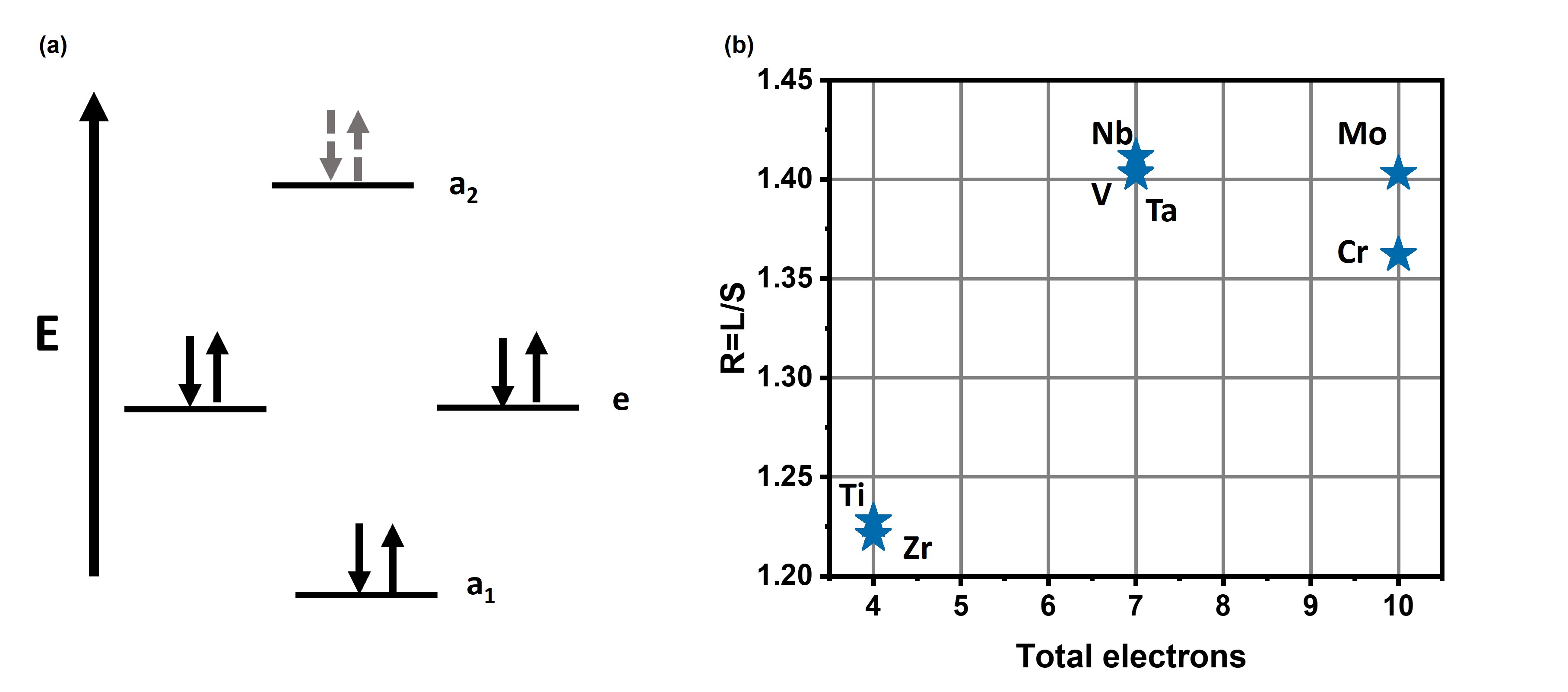}
    \caption{(a) Energy level diagram with 6-8 electrons per M$_3$ cluster favors trimer formation in the crystal system (b) Optimum value of breathing distortion R = 1.41 favors the trimerization.}
    \label{Elabel}
\end{figure}

To identify trends across this family, we analyze the results from the hypothetical compounds listed in Table \ref{tab:compact_table}. Orbital energy level diagrams for representative materials are shown in Figure \ref{Elabel}a, highlighting how trimer valence states influence electronic configurations. Solid black arrows correspond to a formal trimer charge of [M$_3$]$^{6+}$, typical of materials such as Na$_2$Ti$_3$Cl$_8$ and Zr$_3$Cl$_8$, while dotted grey arrows represent [M$_3$]$^{7+/8+}$ states, characteristic of systems like Nb$_3$X$_8$ and Li$_x$ScMo$_3$O$_8$.

To further explore structure–stability relationships, we plot the bond length ratio $R$ (defined as the ratio of long to short M–M bond lengths) against the number of $d$-electrons per M$_3$ cluster for all systems in Table \ref{tab:compact_table} (Figure \ref{Elabel}b). This analysis reveals that 6–8 $d$-electrons per trimer consistently correlates with the formation of strong breathing distortions with a large R=1.41, and resulting molecular orbital formation.

\subsection{Role of Correlation Strength: Effect of Hubbard $U$}

V$_3$Cl$_8$ and Ta$_3$Cl$_8$ meet the key stability criteria identified in the previous section: both have an $R$ ratio near 1.40 and a trimer electron count of seven, corresponding to a [M$_3$]$^{8+}$ valence state—similar to Nb$_3$Cl$_8$, which is known to be stable. However, neither V$_3$Cl$_8$ nor Ta$_3$Cl$_8$ have been experimentally synthesized.

To understand this discrepancy, we examine the role of electron correlation using DFT+$U$. The Hubbard $U$ correction captures on-site Coulomb interactions that are often underestimated in conventional DFT, especially for localized $d$-electrons. By tuning $U$, we assess how correlation strength influences the electronic structure and structural stability of these compounds.

\begin{figure}[h]
    \centering
    \includegraphics[height=0.35\textheight]{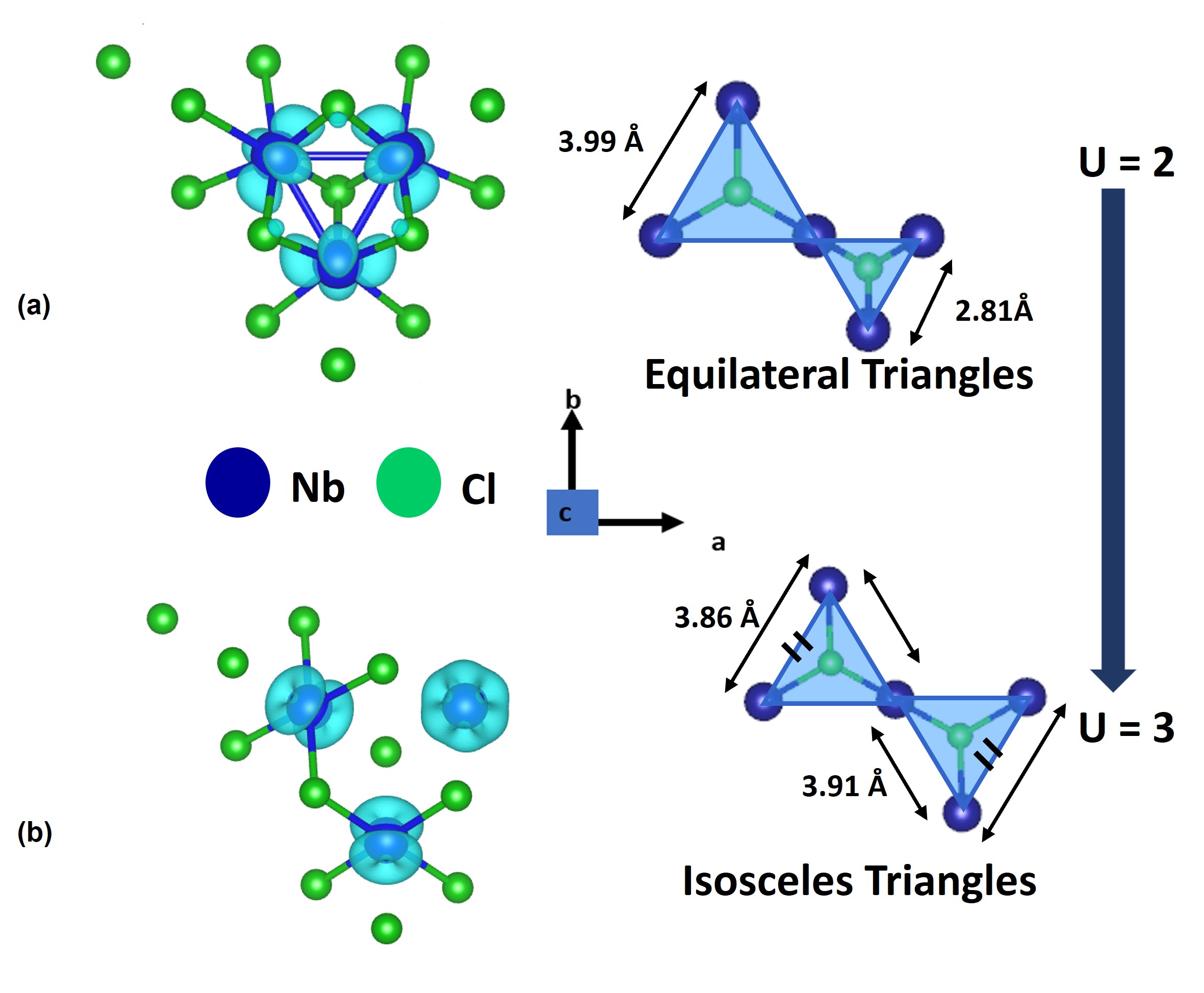}
    \caption{(a) Nb$_3$Cl$_8$ exhibits a breathing distortion under a Hubbard U of 2 eV, which stabilizes a trimerized phase (b) At a Hubbard U of 3 eV, Nb$_3$Cl$_8$ shows no breathing distortion and instead favors a charge-ordered structure.}
    \label{Hubbard}
\end{figure}



We performed DFT+$U$ calculations on Nb$_3$Cl$_8$ using $U$ values ranging from 0 to 3~eV to study the effect of correlation strength on trimer stability. For $U = 0$ to 2~eV, the ground-state structure exhibits trimer formation with clear breathing distortions and spin polarization. In this state, the Nb atoms form triangles with alternating bond lengths of 2.82~\AA\ and 3.98~\AA, as shown in Figure \ref{Hubbard}a.

At $U = 3$~eV, the system undergoes a structural transition to a charge-ordered state, where trimerization is suppressed. The resulting geometry forms isosceles triangles with more uniform bond lengths of 3.86~\AA\ and 3.91~\AA\ (Figure \ref{Hubbard}b). This shift reflects a change in the ground-state character—favoring charge ordering and high-spin configurations at higher correlation strengths.

The effect of the Hubbard $U$ on structural and electronic properties varies significantly across material systems. In some metal–insulator transition compounds, $U$ has been shown to enhance breathing distortions, as previously reported for example in rare earth nickelates \citep{AlexandruB.Georgescu}. In contrast, our calculations reveal that, within this class of materials, an increase in $U$ suppresses breathing distortions and instead favors a charge-ordered ground state. This observation helps explain the experimental absence of V$_3$Cl$_8$. 

\begin{figure}[h]
    \centering
    \includegraphics[height=0.45\textheight]{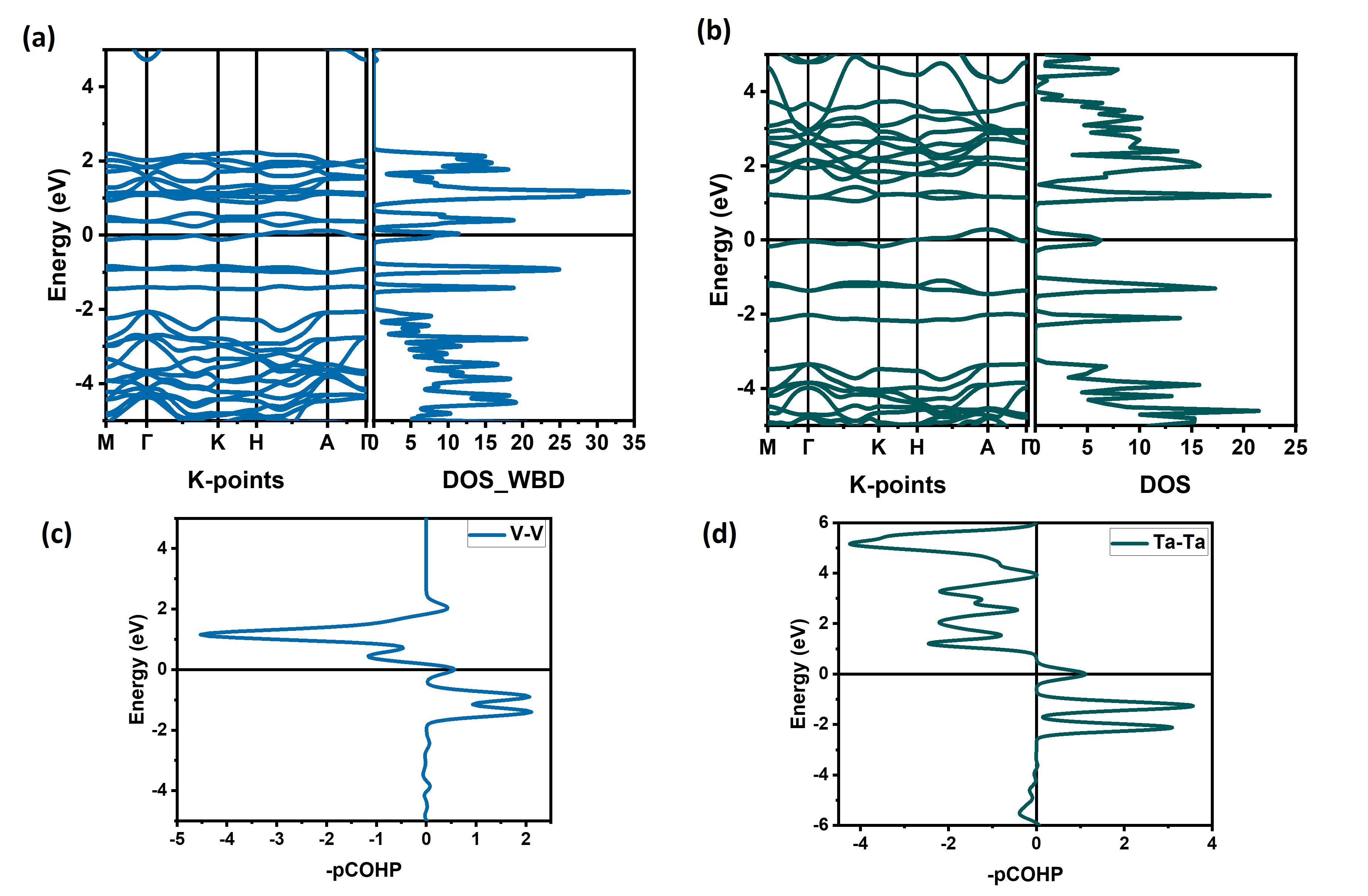}
    \caption{(a) Electronic band structure and DOS of V$_3$Cl$_8$  (b)  Electronic band structure and DOS of Ta$_3$Cl$_8$ (c) COHP plot of V$_3$Cl$_8$ (d) COHP plot of Ta$_3$Cl$_8$.}
   \label{Bands2}
\end{figure}

In contrast, the case of Ta$_3$Cl$_8$ is less clear. The band structure and DOS (SI Figure \ref{Bands2}b), along with the COHP plot (SI Figure \ref{Bands2}d), do not reveal any electronic or bonding instability that would preclude trimer formation. To explore this further, we performed a convex hull analysis to assess the thermodynamic stability of Ta$_3$Cl$_8$. From OQMD \cite{saal2013, kirklin2015} we find that Ta$_3$Cl$_8$ is most likely to decompose into Ta and TaCl$_4$. Our calculations indicate that, indeed, Ta$_3$Cl$_8$ is thermodynamically unstable, tending to decompose into Ta and TaCl$_4$, as described by Equation \ref{Equation}. The energy above the convex hull ($\Delta E$) is calculated to be 8~meV/atom, suggesting that the decomposition is energetically favorable. As the d-orbitals of Ta are more delocalized than those of Nb, its tendency to form metal-metal bonds is even higher: Ta$_3$Cl$_8$ decomposes into a mix of metallic Ta, and TaCl$_4$ (which displays Ta-Ta dimers itself), consistent with our framework as described in Figure \ref{fig:Main}.

\begin{equation}
    \text{Ta}_3\text{Cl}_8 \rightarrow \text{Ta} + 2\text{TaCl}_4
    \label{Equation}
\end{equation}

\section*{Conclusion}

In this work we showed trimer formation in kagome halide systems is driven by a combination of orbital symmetry, electron filling, and electronic correlation: key features of Correlated Electron Molecular Orbital (CEMO) materials. In particular, we find that a trimer electron count of 6–8 ensures optimal occupation of bonding molecular orbitals while minimizing antibonding character, stabilizing structural breathing distortions and supporting correlated insulating behavior - a characteristic shared among a wide range of Kagome materials, from van der Waals kagome halides, to lithium molybdenum oxides which can be electrochemically doped\citep{Wyckoff2022}; and an intermediate correlation strength is required to prevent either the formation of localized states (i.e. charge order) or extended metal-metal bonds.

While this study focuses on zero-temperature behavior, further work is needed to investigate how crystal and electronic entropy influence trimerization, magnetic ordering, and phase stability. Such studies will be essential for guiding the design of robust, high-performance kagome-based materials for future electronic and quantum technologies.

This work provides a general framework for identifying and designing materials displaying molecular orbital states extended across transition metal clusters. By linking symmetry, electron count, and correlation strength with structural and electronic stability, we offer predictive guidelines for engineering next-generation quantum materials with tunable correlated electron and multiferroic phases.

\subsection{Computational Methodology}

All first-principles electronic structure calculations were performed using density functional theory (DFT) and DFT + U (Hubbard U correction), implemented in the Quantum ESPRESSO (QE) package \cite{Paolo, Giannozzi, Baseggio}. We used projector augmented wave (PAW) pseudopotentials, a plane-wave energy cutoff of 544 eV, and a Monkhorst-Pack k-point mesh of 4 × 4 × 3, determined after convergence testing. The Perdew–Burke–Ernzerhof (PBE) exchange-correlation functional was used\citep{perdew1998perdew}.

Calculations were carried out on a formula unit M$_3$X$_8$ unit cell (11 atoms), and all structures were optimized using the Broyden–Fletcher–Goldfarb–Shanno (BFGS) algorithm before density of states (DOS) and band structure calculations.

Crystal Orbital Hamilton Population (COHP) analysis was performed using the LOBSTER package \cite{Richard, Dronskwski, Deringer}, based on wavefunction outputs from QE. Visualization and plotting were carried out using wXDragon and Origin to support the chemical stability analysis of Nb$_3$Cl$_8$. Unless otherwise specified, COHP plots reflect bonding and antibonding orbital occupation within a 3 $\AA$ interaction radius (e.g., for Nb–Nb and Nb–Cl pairs) \cite{Steinberg2018}. All structural visualizations were generated using VESTA \cite{Momma}.

\begin{acknowledgement}
The authors thank Tyrel McQueen, Bipasa Samanta and Madison Genslinger for helpful discussions. This research was supported by Indiana University startup funds, and in part by the Lilly Endowment, Inc., through its support for the Indiana University Pervasive Technology Institute. 
All three authors designed the project; VK and JB performed the calculations, VK wrote the manuscript with input from all authors. ABG supervised the project.

\end{acknowledgement}

\bibliography{achemso-demo}

\clearpage
\subsection{\centering Supporting Information:}
\title{\textbf{Symmetry-Driven Trimer Formation in Kagome Correlated Electron Materials}}

\centering
\author{Varsha Kumari, }
\author{Julia Bauer, }
\author{Alexandru B. Georgescu}

\begin{figure}[h]
    \centering
    \includegraphics[height=0.25\textheight]{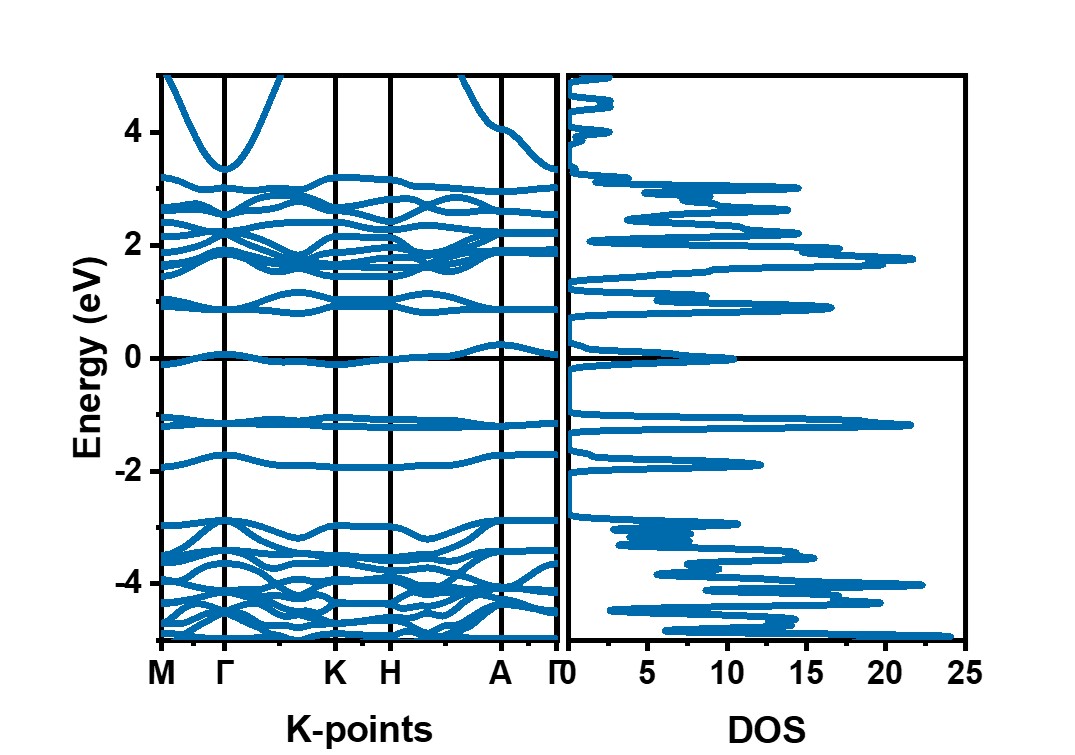}
    \caption{(a) Electronic band structure and density of states (DOS) of Nb$_3$Cl$_8$ obtained from DFT calculation without including spin polarization.}
   \label{fig:enter-label}
\end{figure}

\begin{figure}[h]
    \centering
    \includegraphics[height=0.25\textheight]{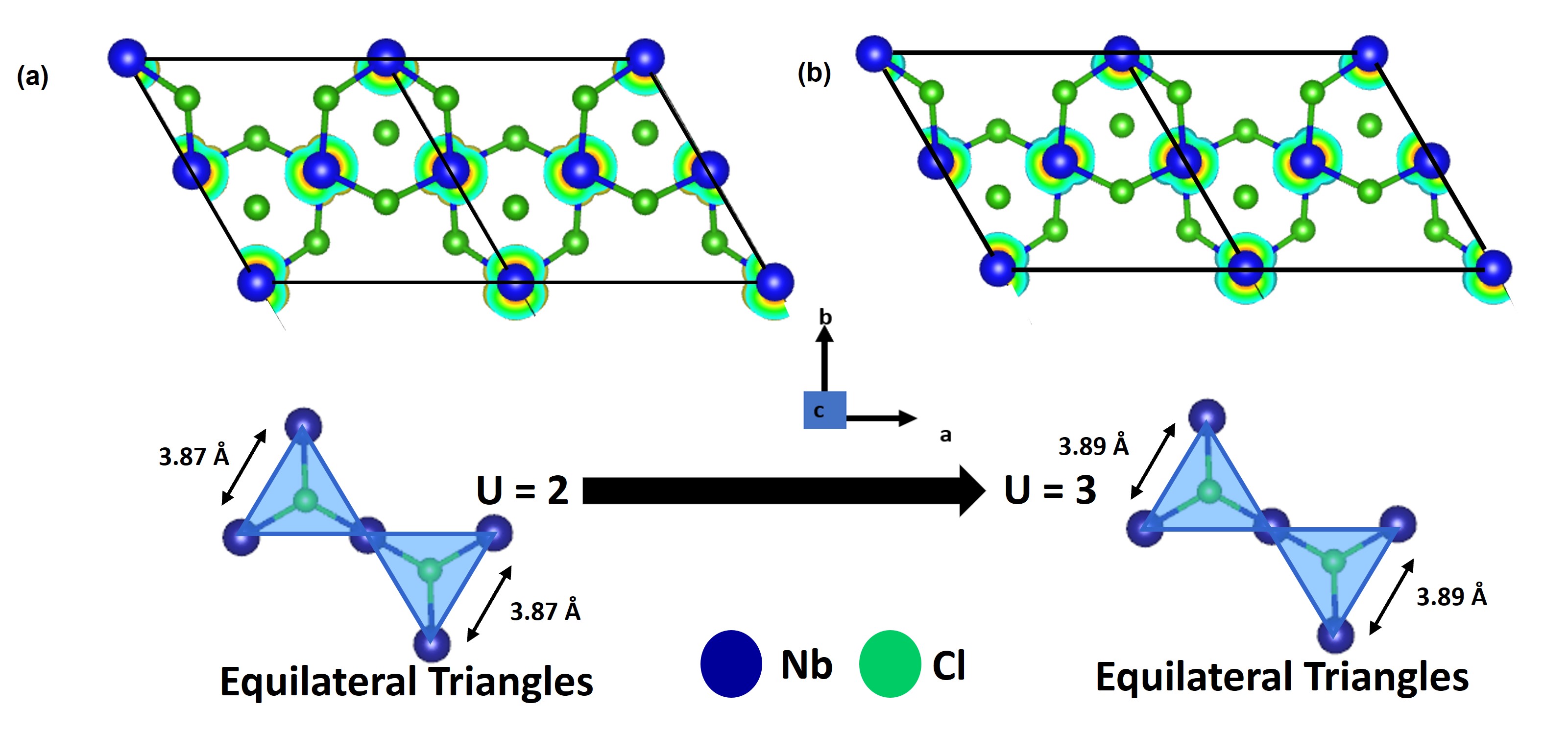}
    \caption{(a) Spin-polarized Nb$_3$Cl$_8$ without breathing distortion with U of 2 eV. (b) Spin-polarized Nb$_3$Cl$_8$ without breathing distortion, with U of 3 eV, showing no change in the structure.}
   \label{Hubbard1}
\end{figure}


\end{document}